\documentclass[twocolumn,showpacs,preprintnumbers,amsmath,amssymb,prl,aps]{revtex4-1}
\usepackage{graphicx}
\usepackage{bm}
\usepackage{float} 
\usepackage{multirow}
\usepackage{amsmath}
\usepackage{gensymb}
\usepackage{epstopdf}
\usepackage[usenames, dvipsnames]{color}

\begin{document}

\title{Evidence of non-trivial Berry phase and Kondo physics in SmBi}
\author{Anup Pradhan Sakhya}
\author{P. L. Paulose}
\author{A. Thamizhavel}
\author{Kalobaran Maiti}
\altaffiliation{Corresponding author: kbmaiti@tifr.res.in}

\affiliation{Department of Condensed Matter Physics and Materials Science, Tata Institute of Fundamental Research, Homi Bhabha Road, Colaba, Mumbai - 400 005, INDIA.}

\begin{abstract}
Realization of semimetals with non-trivial topologies such as Dirac and Weyl semimetals, have provided a boost in the study of these quantum materials. Presence of electron correlation makes the system even more exotic due to enhanced scattering of charge carriers, Kondo screening etc. Here, we studied the electronic properties of single crystalline, SmBi employing varied state of the art bulk measurements. Magnetization data reveals two magnetic transitions; an antiferromagnetic order with a N\'{e}el temperature of $\sim$ 9 K and a second magnetic transition at a lower temperature (= 7 K). The electrical resistivity data shows an upturn typical of a Kondo system and the estimated Kondo temperature is found to be close to the N\'{e}el temperature. High quality of the crystal enabled us to discover signature of quantum oscillation in the magnetization data even at low magnetic field. Using a Landau level fan diagram analysis, a non-trivial Berry phase is identified for a Fermi pocket revealing the topological character in this material. These results demonstrate an unique example of the Fermiology in the antiferromagnetic state and opens up a new paradigm to explore the Dirac fermion physics in correlated topological metal via interplay of Kondo interaction, topological order and magnetism.
\end{abstract}

\date{\today}

\maketitle

\section{I. Introduction}
Topological insulators (TIs) \cite{Hasanreview} having bulk band gaps and robust symmetry protected surface states have attracted enormous interest in the recent time due to their importance in exploring rich fundamental science and application in advanced technology. While impurity on the surface of these materials often exhibit interesting evolutions \cite{Deep1-Bi2Se3,Deep2-Bi2Se3Deep-Bi2Se3}, breaking of suitable symmetry helped to realize semimetals with non-trivial topologies such as Dirac \cite{Liu} and Weyl semimetals, \cite{Burkov}. The electronic structure of Dirac or Weyl semimetals are gapless and contain Dirac or Weyl points; the crossing of the linearly dispersive bands \cite{Liu, Burkov}. These materials provide an wonderful opportunity to understand the physics of relativistic particles in low-energy condensed matter systems. Unique bulk band topology in these semimetals exhibit exotic physical properties such as extreme magneto-resistance (XMR), high charge carrier mobility, potential topological superconductivity, and chiral-anomaly driven negative magnetoresistance (MR) \cite{Binghai}. Till date most of the topological materials discovered experimentally are weakly correlated systems, where the effective single particle descriptions adequately captures the electronic properties.

Strong spin-orbit coupling (SOC) in the correlated electron systems may lead to topologically protected surface states in addition to other interesting properties \cite{KL1-Swapnil,KL2-Swapnil}. Interestingly, in the correlated topological insulators, the electron correlation enhances the effective electron mass leading to insulating bulk ground state while the surface states behave like Dirac Fermions despite the presence of strong electron correlation. For example, a Kondo insulator, SmB$_6$, has been predicted to host topological surface states within the Kondo gap \cite{SmB,Dzero,Kim}. Subsequent studies, however, showed anomalies such as quantum oscillations \cite{SmB,Tan} and finite Fermi surface \cite{Mark-ARPES,Sakhya-AIPProc,Sakhya-SREP}. RSb/RBi (where R = rare earth) family of materials are also proposed to be potential candidates for correlated topological semimetals \cite{Tafti,Ding,MZeng}. Based on the band structure calculations LaX (X = P, As, Sb, Bi) have been classified as $Z_2$-topological metals \cite{MZeng}. Angle resolved photoemission spectroscopy (ARPES) on LaSb indicate that the material is topologically trivial and its properties can be well explained by electron-hole compensation \cite{Ding}. On the other hand, LaBi is toplogically nontrivial exhibiting three Dirac cones; one Dirac cone at the surface Brillouin zone (SBZ) center and two are at the SBZ corner \cite{Nayak}. Another ARPES study contradicted these results suggesting that the surface states in LaBi are highly unusual possessing the parabolic bottom band and linear top band \cite{wu}. The origin of such unusual mass acquisition is yet to be understood \cite{wu}.

In this class of materials, SmBi exhibits band inversion between the Sm 5$d$ and Bi 6$p$ states similar to LaBi \cite{cao} although the ARPES study by Li \textit{et al.} could not find the predicted surface state at the $\Gamma$ point presumably due to its strong hybridization with the bulk states. They found two surface states at the high symmetry point, $M$, which interact with each other and yield a peculiar pair of gapped trivial surface states \cite{Peng}. ARPES and transport studies of (Pr, Sm)Sb and (Pr, Sm)Bi revealed signature of extreme magnetoresistance (XMR) due to good electron-hole compensation \cite{Zhongzheng}. Evidently, presence of strong electron correlation leads to unusual complexity in their properties. Multiple bands cross the Fermi level making the identification of the Fermi surfaces and its behavior difficult.

In order to address the issues, we prepared very high quality single crystals of SmBi and employed de Haas van Alphen (dHvA) measurements to derive the Fermiology, which relies on quantum oscillations as a function of external magnetic field. The phase offset of the quantum oscillation is related to Berry phase associated with the cyclotron orbits and provides a means to experimentally access the signature of the Dirac cone \cite{Wright}. We discover pronounced quantum oscillations even at low magnetic field; the analysis of the experimental data reveals presence of multiple Fermi pockets and a signature of a non-trivial Berry phase. Resistivity upturn at around $\sim$ 12 K typical of a Kondo behavior and the signature of Berry phase provide evidence of Dirac fermion physics for the electronic properties of this correlated material.

\section{II. Experimental and Crystal Structure}

High qualtiy single crystals of SmBi was grown following flux method with high purity Sm (99.9\%), Bi (99.998\%) and In (99.9999\%) [molar ratio of Sm:Bi:In was 1:1:10]. The mixture was kept in a recrystallized alumina crucible and sealed in an evacuated quartz ampoule at a vacuum of 10$^{-6}$ Torr. The sealed ampoule heated to 1050 $^o$C  at a rate of 60 $^o$C/hr, kept for 24 hr and then the temperature was reduced very slowly (2 $^o$C/hr) to 700 $^o$C. The ampoule was centrifuged at 700 $^o$C to remove the flux. We have verified the crystal structure using powder $x$-ray diffraction method. Energy dispersive analysis of $x$-rays at various regions of the single crystal established homogeneity and stoichiometry of the material. The orientation and single crystallinity of the sample is determined using Laue diffraction. The electrical resistivity was measured down to 5 K using a homemade setup. The dc magnetic susceptibility and the magnetization measurements of SmBi were performed in the temperature range, 1.8 K to 300 K using a superconducting quantum interference device (SQUID). The dHvA oscillations were probed through magnetization measurements in a vibrating sample magnetometer from Quantum Design. SmBi forms in rock salt structure as shown in Figure \ref{figure1}(a). We prepared single crystals of SmBi starting with highly pure ingredients and characterized the sample using varied methods. A typical Laue diffraction pattern shown in Figure \ref{figure1}(b), exhibit clear four-fold rotational symmetry pattern corresponding to (001) plane of the cubic structure with well defined spots that demonstrates high quality of the crystals.

\section{III. Results and Discussion}
\begin{figure}
\includegraphics[scale=0.4]{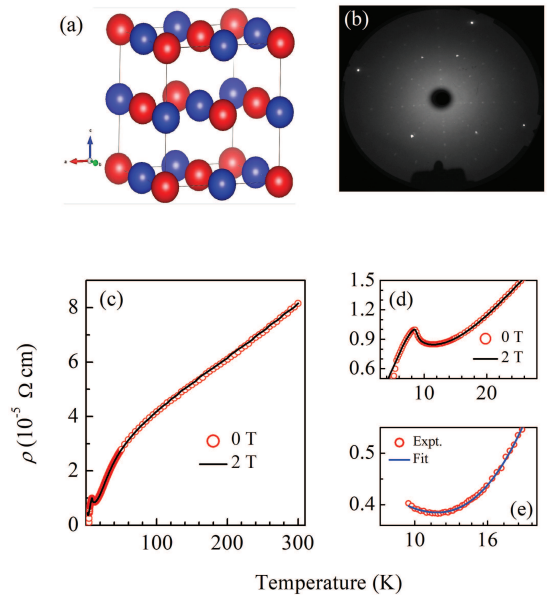}
\vspace{-14ex}
\caption{(color online) (a) Crystal structure of SmBi. The red and blue colored solid spheres denote the Sm and Bi atoms respectively. (b) Laue diffraction pattern of SmBi (001) single crystal. (c) Temperature dependent electrical resistivity measured in the presence of a field of 0 Tesla and 2 Tesla. The data show an upturn at about 12 K and then a sharp decrease below 9 K. (d) Enlarged view of the electrical resistivity in the low temperature regime. (e) The low-temperature upturn in resistivity versus $ln T$. The solid line shows the fit over the temperature range from 9.4 K to 18 K using equation \ref{RT}.}
\label{figure1}
\end{figure}

In Figure \ref{figure1}(c), we show the electrical resistivity, $\rho$  as a function of temperature for electric current parallel to the [100] direction. A typical metallic behavior is observed with the decrease in temperature from room temperature until around 12 K along with an upturn at lower temperatures till 9 K and then there is a sharp decrease as shown in an expanded temperature scale in Figure \ref{figure1}(d). A broad hump observed at about 50 K may be due to the crystal field effect as found in other systems \cite{thamiz}. The resistivity upturn cannot be captured using typical behavior due to disorder and/or other localization effects at low temperatures. In order to investigate if Kondo interactions plays a role here, the resistivity upturn in this narrow temperature range is simulated using Hamann's model \cite{liang, hamann}, where the typical Kondo scattering is captured by the $ln T$ dependence of resistivity. In this model, the contribution due to the scattering by the magnetic impurity, $\rho_H$ can be expressed as:
\begin{equation}
\label{Rd}
\rho_H = \frac{\rho_u}{2} (1-\frac{ln(T/T_K)} {\sqrt{ln^2 (T/T_K) + S(S+1) \pi^2}})
\end{equation}
where, ${\rho_u}$ is the resistivity at the unitarity limit, $T_K$ is the Kondo temperature and $S$ is the spin moment of the magnetic impurity. Above the resistivity minimum, the major contribution to resistivity arises from electron-phonon scattering, which can be captured by fitting the resistivity, ${\rho(T)}$ in the higher temperature range using an expression, $a + bT^n$. From this fitting, we found $b$ = 1.6$\times10^{-7}$ and $n$ = 1.3. The resistivity in the low temperature range (9.4 K to 18 K) is then fitted using an expression that includes the Hamann term, along with the electron-phonon contribution $bT^n$ and a temperature-independent term ${\rho_0}$,
\begin{equation}
\label{RT}
\rho_{T} = \rho_H + bT^{n} + \rho_0
\end{equation}
The fit to the experimental data is shown in Figure \ref{figure1} (e) for the parameter values; $\rho_u$  =  6.54$\times10^{-6}$ $\ohm$, $\rho_0 = 2.8\times10^{-6} \ohm$, S =0.02 and $T_K$ = 9.2 K. Good representation of the experimental data indicates importance of Kondo coupling of the Sm moments with the conduction electrons in its electrical transport properties leading to resistivity upturn.

\begin{figure}[h]
\centering
\includegraphics[scale=0.4]{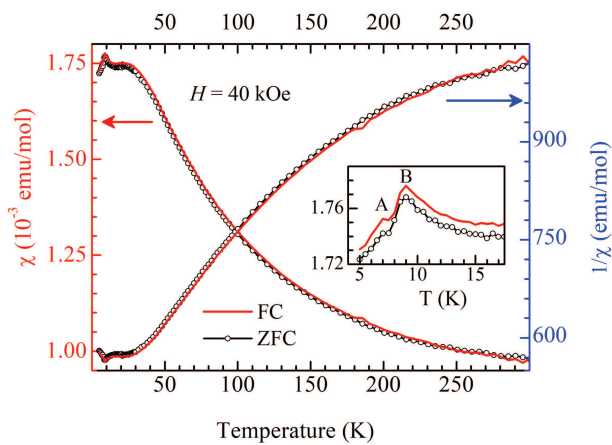}
\vspace{-24ex}
\caption{(color online) Zero field cooled and field cooled magnetic susceptibility, $\chi$, vs temperature plot (left axis) and inverse magnetic susceptibility, 1/$\chi$, vs temperature plot (right axis) measured under magnetic field of $B$ = 40 $k$Oe. Inset shows the enlarged plot of $\chi$ at T$<$ 18 K. The peaks labeled A and B in the inset shows signatures of the two magnetic transitions.}
\label{figure2}
\end{figure}

Below 9 K, the resistivity shows a sharp decrease as often observed in various rare-earth intermetallics due to antiferromagnetic ordering via Ruderman-Kittel-Kasuya-Yosida (RKKY) interactions between the uncompensated part of the magnetic moments \cite{kastner}. This is verified by the magnetic susceptibility measurements. The temperature-dependent molar magnetic susceptibility $\chi(T) = M/H$ measured at the magnetic field,  $B$ = 40 $k$Oe and it's inverse, $1/\chi(T)$ are shown in Figure \ref{figure2}. A dome-shaped curve centered at about 25 K is evident in $\chi$(T) plot akin to the observations in the Kondo insulator, SmB$_6$ \cite{Fisk}. With the decrease in temperature, $\chi$(T) increases as expected in a paramagnetic system and shows a sharp peak at around 9 K (labelled as B) \cite{Hulliger}. Interestingly, the experimental data reveal a second peak (labelled A) at 7 K (see the inset of the Figure \ref{figure2}) indicating presence of additional magnetic transition in the system. Further experimental studies, especially neutron diffraction measurement are necessary to understand this complex behavior.

\begin{figure}[h]
\includegraphics[scale=0.4]{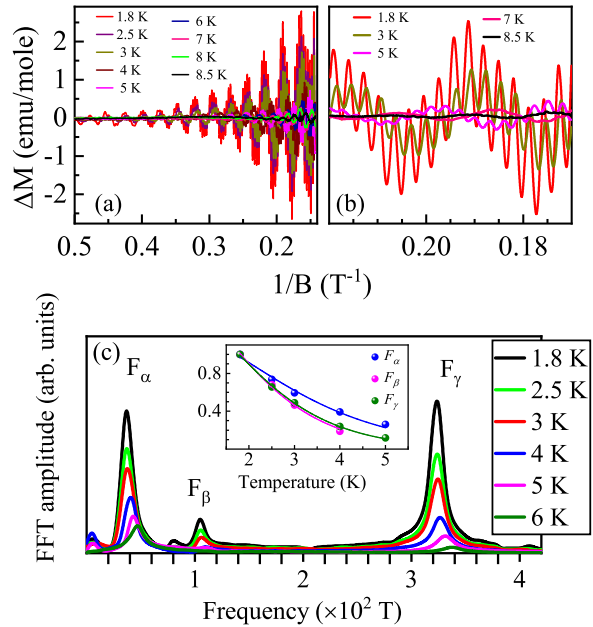}
\vspace{-12ex}
\caption{(color online) (a) dHVa oscillations obtained by subtracting polynomial background from magnetic measurement, plotted with inverse magnetic field (1/B) at different temperatures. (b) dHVa oscillations at some selected temperatures in an expanded $x$-scale. (c) Fast fourier transform of the dHVa oscillations exhibit three principal frequencies $F_{\alpha}$ at 37 T, $F_{\beta}$ at 105 T and $F_{\gamma}$ at 324 T. Inset in (c) -  FFT amplitude of the dHVa oscillations as a function of temperature for the frequencies $F_{\alpha}$, $F_{\beta}$ and $F_{\gamma}$ (symbols). Solid lines are the fits using the thermal damping factor of the Lifshitz-Kosevich formula.}
\label{figure3}
\end{figure}

We have measured the magnetization as a function of magnetic field at different temperatures. The magnetization isotherms of SmBi shows a paramagnetic type field dependence. In addition, we observe significant oscillation in the magnetization as a function of the applied field although the magnetic field applied is small compared to the kind of field required to get discernible oscillations. In order to extract the oscillations, the background magnetization curve, $M_0(B)$ has been simulated using a polynomial; a fifth order polynomial replicates the background well. The subtracted magnetization, $\Delta M$ (= $M(T) - M_0(B)$) as a function of the applied magnetic field, $B$ ($\parallel c$) is shown in Figure \ref{figure3}(a) exhibiting clear signature of de Haas van Alphen (dHvA) oscillations is observed. The dHvA oscillations for some selected temperatures in an expanded magnetic field scale are shown in Figure \ref{figure3} (b) for clarity. Interestingly, the oscillations at 1.8 K can be observed down to a very low magnetic field of about 1.1 T. With the increase in temperature, the oscillations dies down drastically and almost disappears above 8.5 K.

The frequency spectra obtained by fast Fourier transform (FFT) of the dHvA oscillations is shown in Figure \ref{figure3}(c). There are three principal frequencies $F_\alpha$ at 37 T, $F_\beta$ at 105 T and $F_\gamma$ at 324 T indicating presence of three Fermi pockets for $B \parallel c$ in SmBi. Signatures of higher harmonics are also observed, which are not shown here for clarity. With the increase in temperature, the amplitude of all the frequencies decreases fast due to the thermal broadening of the Landau levels.

\begin{figure}[h]
\includegraphics[scale=0.4]{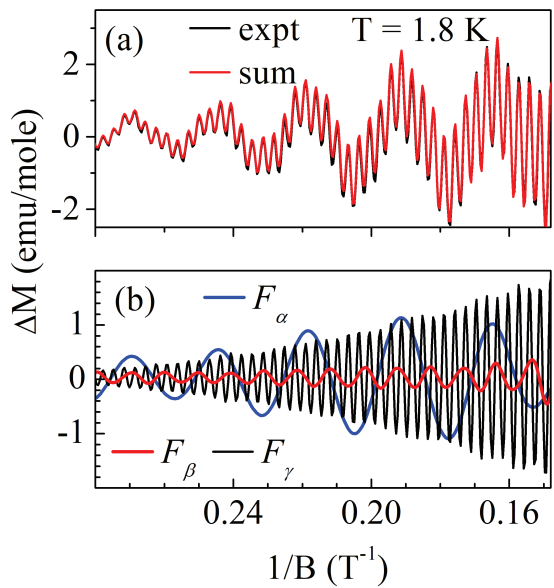}
\vspace{-18ex}
\caption{(color online) (a) Oscillatory part of magnetization as a function of 1/B at 1.8 K (the black line is the experimental data and the red line is the sum of the three filtered frequencies). (b) The three filtered oscillatory parts of magnetization corresponding to $F_\alpha$ (blue line), $F_\beta$ (thick red line) and $F_\gamma$ (black line).}
\label{figure4}
\end{figure}

These three pockets can be filtered from the total oscillation data by using a Band-pass filter. The filtered oscillatory parts for the $F_\alpha$ (blue line), $F_\beta$ (red line) and $F_\gamma$ (black line) branch are shown in Figure \ref{figure4} (b). The sum of the three filtered frequencies (red line) in Figure \ref{figure4}(a) matches very well both in amplitudes and phases with the experimental data (black line) providing confidence on the analysis.

\begin{table}
  \centering
  \caption{Fermi surface parameters for SmBi obtained from the dHvA oscillations data.}\label{Table1}
  \begin{ruledtabular}
  \begin{tabular}{l|c|c|c|c}
  F(T) & $A_F$ (10$^{-2}$\AA$^{-2}$) & $k_F$ (10$^{-2}$\AA$^{-1}$) & $m^*/m_e$ & $v_F$ (10$^5$ m/sec)\\
  \hline
  37 & 0.354 & 3 & 0.216 & 1.6 \\
  105 & 1.01 & 5.65 & 0.304 & 2.1\\
  324 & 3.1 & 9.9 & 0.287 & 3.98\\
  \end{tabular}
  \end{ruledtabular}
  \vspace{-2ex}
 \end{table}

The above results suggest that there are three Fermi pockets having finite cross section perpendicular to the $c$ axis. The frequency (in inverse field) of the oscillations is related to the $k$-space cross-sectional area of the Fermi surface $A_F$ by the Onsager relation, $F =(\phi_0/2\pi^2) A_F$ (magnetic flux quantum, $\phi_0 = h/2e$). Using Onsager relation, the Fermi surface cross-section ($A_F$), corresponding Fermi momentum ($k_F$), and Fermi velocity ($v_F$) are calculated and listed in the Table \ref{Table1}. Temperature dependence of the oscillation amplitudes are shown in the inset of Fig. \ref{figure3}(c), which is calculated using the thermal damping factor in the Lifshitz-Kosevich (LK) formula \cite{shoenberg},
\begin{equation}
\label{LK}
R_T = (2\pi^2 k_B T / \beta) /sinh(2\pi^2 k_BT / \beta)
\end{equation}
where $\beta = e\hbar B/m^*$, $m^*$ is the cyclotron effective mass. Due to the presence of multiple branches in the quantum oscillation, the fitting parameter $B$ has been defined as the average inverse field of the FFT interval, and the amplitudes of the FFT peaks are used for fitting. The best fit is shown by solid line in the inset of Fig. \ref{figure3}(c). Using the fitting parameters, we have also calculated the cyclotron mass, $m^*$ for the $F_\alpha$, $F_\beta$ and the $F_\gamma$ pockets and are listed in Table \ref{Table1}.

\begin{figure}
\vspace{-3ex}
\includegraphics[scale=0.4]{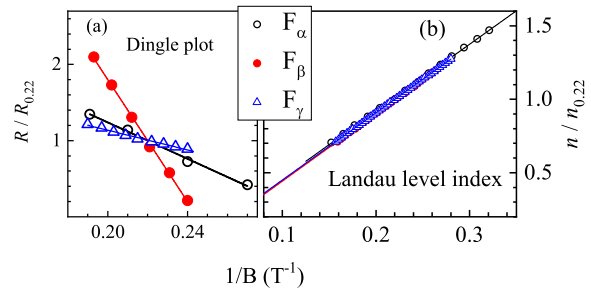}
\vspace{-35ex}
\caption{(color online) (a) The Dingle plots for the frequencies $F_\alpha$ (open circle), $F_\beta$ (closed circle) and $F_\gamma$ (open triangle) along with the fits using Lifshitz-Kosevich formula. In the $y$-axis, $R = ln [\Delta M B^{1/2} sinh(14.69m^* T/B)]$. The magnitude of $R$ is normalized by it's value at $1/B$  = 0.22 T$^{-1}$ [1.045, 0.1667 and 1.65 for $F_\alpha$, $F_\beta$ and $F_\gamma$, respectively]. (b) Landau fan diagram extracted from the quantum oscillation data; the normalized Landau band index, $n/n_{0.22}$ vs the reciprocal of the field, $1/B$ [$n$ = Landau band index and $n_{0.22}$ is the value of $n$ at $1/B$ = 0.22 T$^{-1}$, which are 8.5, 22.5 and 71 for $F_\alpha$, $F_\beta$ and $F_\gamma$, respectively]. Universality in field dependence for all the Fermi pockets.}
\label{figure5}
\end{figure}

The Dingle temperature has been calculated by fitting the dHvA oscillation amplitudes with inverse magnetic field to the Dingle damping term,
\begin{equation}
\label{RD}
R_D =\exp(-2\pi^2 k_B m^* T_D/\hbar eB)
\end{equation}
To improve accuracy in fittings, we have separated the oscillation components of $F_\alpha$, $F_\beta$ and $F_\gamma$ bands by filtering out irrelevant oscillations, and then extracted the oscillation amplitude of 1.8 K as a function of $1/B$. The best fit using the LK equation is shown in Fig. \ref{figure5} (a) and the estimated Dingle temperatures, $T_D$ are 3.9 K, 1.5 K and 2.5 K for  $F_\alpha$, $F_\beta$ and the $F_\gamma$ , respectively. The small Dingle temperature indicates little influence from disorder due to impurities. Using the estimated Dingle temperatures, the mobility, $\mu_q = (e\hbar/2\pi k_B m^* T_D)$ of the charge carriers corresponding to the three Fermi surfaces has been found to be 2510 cm$^2$/Vsec, 4650 cm$^2$/Vsec and 2954 cm$^2$/Vsec, respectively.

We now investigate the possible topological nature of the bands by constructing the Landau fan diagram. The Landau level assignment is done using the peak positions of the oscillation component of each frequency shown in Fig. \ref{figure3} (c). When a closed cyclotron orbit is quantized under an external magnetic field $B$, then according to the Lifshitz-Onsager (LO) quantization rule,
\begin{equation}
\label{LO}
\frac{A_n\hbar}{eB} = 2\pi(n + \gamma - \delta)
\end{equation}
Here, $A_n$ is the Fermi surface cross section of the Landau level (LL) $n$, $\gamma = \frac{1}{2}-\frac{\phi_B}{2\pi}$ is the phase factor ($\phi_B$ is the Berry phase) and $\delta$ is an additional phase shift ($\delta$ = 0 in two dimensions, = $\pm$1/8 in three dimensions) \cite{JCao,Yupeng}. Clearly, a vanishingly small $\gamma$ makes $\phi_B$ close to $\pi$, thereby provide a signature of non-trivial Berry phase. The Landau fan diagram, a plot of Landau index number $n$ ($y$-axis) vs $1/B$ ($x$-axis), is shown in Fig. \ref{figure5}(b). Here, the Landau level index, $n$ ($n$ + 1/2) is assigned to each $\Delta M(B)$ minimum(maximum) \cite{JCao} and is calculated using LO formula as shown in the figure by lines superimposed over the data points (symbols). Interestingly, a normalization of $n$ by its values at $1/B$ = 2.2 puts all the data on a single straight line indicating an universal behavior of the charge carriers in all the Fermi pockets.

The magnitude of $|\gamma-\delta|$ for the $F_\alpha$, $F_\beta$ and $F_\gamma$  pockets are found to be 0.07,  0.34 and 0.4, respectively. The small value of $|\gamma-\delta|$ for the $F_\alpha$ pocket (within $\pm$1/8) reveals an evidence of a non-trivial Berry phase of $\pi$ and indicates its correspondence to Dirac fermions \cite{Yupeng,Novoselov}.

\section{IV. Conclusion}

From the above results it is clear that very high quality single crystal of SmBi helped to unveil several puzzles. Transport measurements exhibit a resistivity upturn at about 12 K with a $lnT$ dependence typical of a Kondo lattice system. The estimated Kondo temperature is found to be 9.2 K. Magnetization data reveals signature of two magnetic transitions; one at 9 K and the other at 7 K. The first transition at 9 K correspond to antiferromagnetic ordering, signature of which is also seen in the resistivity data via a sharp decrease in resistivity as often observed in various strongly correlated systems \cite{KL2-Swapnil}. The N\'{e}el temperature of 9 K is close to the estimated Kondo temperature suggesting a delicate competition between the Kondo effect and the indirect exchange interaction. Clearly the compensation of magnetic moment is far from complete providing a possible case of spin-density wave quantum criticality \cite{Millis,SDW_Swapnil,gegenwart}. The origin of second transition is not clear yet. However, it's striking similarity with other Kondo lattice systems such as CeB$_6$ is interesting.

While doing the magnetization studies at low temperatures, we discover beautiful quantum oscillations at a field as low as 1.1 Tesla. Prominent dHvA quantum oscillations are observed at temperatures upto about 8.5 K where the system is in antiferromagnetic phase. From the analysis of the magnetization data, three intriguing oscillation branches, $F_\alpha$, $F_\beta$, and $F_\gamma$ corresponding to three Fermi surfaces are observed. The Berry phase was unambiguously characterized by extrapolating the observed dHvA oscillations to the zero-energy Landau level using Landau fan diagram. Signature of a non-trivial Berry phase is discovered for the $F_\alpha$ Fermi-pocket, suggesting the nontrivial topological character. Ironically, although the system possesses Kondo-type interactions, the effective mass for all the Fermi pockets found is very small along with high mobility. These results thus provide an unique example of a system exhibiting Kondo interactions along with highly diminished effective mass and an evidence for a nontrivial Berry phase, which provides an opportunity for studying the interplay between electron correlations, topology and magnetism.

\section{Acknowledgments}
Authors acknowledges financial support from the Department of Atomic Energy, Govt. of India under the project no. 12-R\&D-TFR-5.10-0100. KM acknowledges financial support from DAE, Govt. of India under the DAE-SRC-OI Award program, and Department of Science and Technology, Govt. of India under J. C. Bose Fellowship program.

\end{document}